\documentstyle[11pt,epsfig,amssymb,amsmath]{article}
\begin{document}
\title{Bounding $f(R,T)$ gravity by particle creation }
\author{$K.\;\;  Asadiyan^{1}$\thanks{Email: kasadyan@gmail.com}\;\;, $ F.\;\; Darabi^{2}$\thanks{Email: f.darabi@azaruniv.ac.ir}\;\;, $P.\;\; Pedram^{1}$\thanks{Email: p.pedram@srbiau.ir}\;\; and $ K.\;\; Atazadeh^{2}$  \thanks{Email: atazadeh@azaruniv.ac.ir}\\$ ^{1} {\small Department\;\; of\;\; Physics,\;\; Science\;\; and \;\;Research\;\; Branch,}$\\${ \;\;Islamic\;\; Azad\;\; University,\;\; Tehran,\;\; Iran}$\\
$^{2} {\small Department\;\; of\;\; Physics,\;\; Azarbaijan\;\; Shahid\;\; Madani\;\; University,}$\\${ Tabriz\;\; 53714-161,\;\; Iran.}$}

\date{\today}
\maketitle
\begin{abstract}
We consider the possibility of the quantum vacuum states in $f(R,T)$ gravity. Particularly, we study the Bogoliubov
transformations associated to different vacuum states for some $f(R,T)$ models. The method consists
of fixing the $f(R,T)$ free parameters by requiring the Bogoliubov coefficients to be minimized. In such a way, the particle production is related to the value of the Hubble parameter and also the given $f(R,T)$ model.
\end{abstract}

\section{Introduction}
  \qquad Recent observational data, such as supernova Ia (SN Ia), Cosmic Microwave Background Radiation (CMBR), Large Scale Structure (LSS)
 and Baryon acoustic Oscillation (BAO) indicated that our universe is expanding at an acceleration rate \cite{perlm,benne,tegma}.
  These observations seem to change the entire picture of our matter filled universe.
  It is now believed that most part of the universe contains dark matter and dark energy \cite{sahni,bamba,copland,amendo,fremen}.
  It is well known that general relativity (GR) based on the Einstein--Hilbert action can not explain
 the acceleration and expansion of the universe. Thus at large scale the Einstein gravity model of general relativity
 break down and a more general action describes the gravitational field.
 
  There are two approaches to explain the modification of Einstein's GR.
  According to the first approach we modify the right hand side of the Einstein equation, {\it i.e.} in the matter sector of the universe.
 Some of the popular ones worth mentioning are Chaplygin gas models \cite{moschella,gorini}, Quintessence scaler field \cite{ratra}, Phantom energy field \cite{gold}, {\it etc}.
 The other approach is based  on modification of the Einstein--Hilbert action, {\it i.e.} in the geometry sector of the universe, giving birth to modified gravity theories.
 Some of the popular models of modified gravity that came into existence in recent times are $f(R)$ gravity \cite{nojiri}, $f(T)$ gravity \cite{ferraro},
 Gauss-Bonnet gravity \cite{carr}, {\it etc}.

In this work we will consider the particle production phenomenon in the context of $f(R,T)$  model as the modification of Einstein Lagrangian by introducing an arbitrary function of scalar curvature $R$ and
 trace of the energy--momentum tensor $T$ \cite{harco}.
 The dependence of $T$ may be introduced by exotic imperfect fluids or quantum effects (conformal anomaly).
 The paper is organized as follow: In section 2, a brief review of modified $f(R,T)$ gravity theory and its field
 equations are presented.
 In section 3, particle production phenomenon in the framework of non--minimally coupled theories of gravity is investigated.
 In section 4, we minimize the Bogoliubov coefficients. In doing so,  we constrain the free parameters of a given $f(R,T)$ model.
 Finally in section 5, we present the conclusion.

\section{A brief review of $f(R,T)$ gravity}
The $f(R,T)$ gravity is a modified theory of gravity, in which Ricci scalar $R$ in the Einstein--Hilbert Lagrangian is
 replaced by the arbitrary function of $R$ and $T$ (trace of energy--momentum tensor $T_{\mu\nu}$).
The action for $f(R,T)$ gravity in the gravitational unit $8\pi G=c=1$ is given by
 \begin{eqnarray}\label{1}
 S=\frac{1}{2}\int\sqrt{-g}\left[f(R,T)+2{\cal L}_{m}\right]\mathrm{d}^4x,
 \end{eqnarray}
 where $g$ is the determinant of metric tensor $g_{\mu\nu}$ and $ {\cal L}_{m}$ describes the matter Lagrangian density.
 The energy--momentum tensor $T_{\mu\nu}$, is defined as
\begin{eqnarray}\label{2}
 T_{\mu\nu}=-\frac{2}{\sqrt{-g}}\frac{\delta(\sqrt{-g}{\cal L}_{m})}{\delta {g^{\mu\nu}}}.
\end{eqnarray}
We assume that the matter lagrangian density depends only on the metric tensor components $g_{\mu\nu}$ so that
\begin{eqnarray}\label{3}
T_{\mu\nu}=g_{\mu\nu}{\cal L}_{m}-2\frac{\delta{\cal L}_{m}}{\delta {g^{\mu\nu}}}.
\end{eqnarray}
By taking the variation of the action (\ref{1}) with respect to the metric tensor, we get the field equation for the $f(R,T)$ gravity as follows
\begin{eqnarray}\label{4}
f_R(R,T)R_{\mu\nu}-\frac{1}{2}f(R,T)g_{\mu\nu}+(g_{\mu\nu}\square-\triangledown_{\mu}\triangledown_{\nu})f_R(R,T)=T_{\mu\nu}-
f_T(R,T)T_{\mu\nu}-f_T(R,T)\Theta_{\mu\nu},
\end{eqnarray}
where $f_R(R,T)$, $f_T(R,T)$ denote the derivatives of $f(R,T)$ with respect to $R$ and $T$, respectively.
Here, $\triangledown_{\mu}$ is the covariant derivative and $\square\equiv\triangledown_{\mu}\triangledown^{\mu}$
 is the d'Alembert operator and $\Theta_{\mu\nu}$ is defined by
\begin{equation}\label{5}
\Theta_{\mu\nu}=g^{\alpha\beta}\frac{\delta {T_{\alpha\beta}}}{\delta {g^{\alpha\beta}}}.
\end{equation}
Using (\ref{3}) in (\ref{5}), we obtain
\begin{eqnarray}\label{6}
T_{\mu\nu}=-2T_{\mu\nu}+g_{\mu\nu}{\cal L}_{m}-2g^{\alpha\beta}\frac{\partial^2{\cal L}_{m}}{\partial g^{\mu\nu}\partial g^{\alpha\beta}}.
\end{eqnarray}
The energy--momentum tensor of the matter is given by
\begin{equation}\label{7}
T_{\mu\nu}=(\rho_{m}+p_m)u_{\mu}u_{\nu}-p_mg_{\mu\nu},
\end{equation}
where $\rho_m$, $p_m$ and $u_m$ are respectively the energy density, the pressure and the four--velocity of a perfect fluid.
The matter Lagrangian is given by ${\cal L}_{m}=-p_m$, so that we have
\begin{eqnarray}\label{8}
\Theta_{\mu\nu}=-2T_{\mu\nu}-p_mg_{\mu\nu}.
\end{eqnarray}
Using this in the field equation (\ref{4}) gives
\begin{eqnarray}\label{9}
R_{\mu\nu}f_R(R,T)-\frac{1}{2}g_{\mu\nu}f(R,T)-(\triangledown_{\mu}\triangledown_{\nu}-g_{\mu\nu}\square)f_R(R,T)=
T_{\mu\nu}+f_T(R,T)T_{\mu\nu}+p_mf_T(R,T)g_{\mu\nu}.
\end{eqnarray}
Now, we rewrite this equation to obtain the Einstein--like equation as
\begin{eqnarray}\label{10}
G_{\mu\nu}=T_{\mu\nu}^{(ef\mbox{}f)}+T_{\mu\nu}^{(curv)},
\end{eqnarray}
where $G_{\mu\nu}$ is the Einstein tensor, $T_{\mu\nu}^{(ef\mbox{}f)}$ is
the \emph{effective} matter energy tensor given by
\begin{eqnarray}\label{11}
T_{\mu\nu}^{(ef\mbox{}f)}=\frac{1}{f_R(R,T)}\left[1+f_T(R,T)T_{\mu\nu}+p_mf_T(R,T)g_{\mu\nu}\right],
\end{eqnarray}
and $T_{\mu\nu}^{(curv)}$ is the energy--momentum tensor corresponding to
the
\emph{curvature} given by
\begin{eqnarray}\label{12}
T_{\mu\nu}^{(curv)}=\frac{1}{f_R(R,T)}\left[\frac{1}{2}g_{\mu\nu}\big(f(R,T)-Rf_R(R,T)\big)+\big(\triangledown_{\mu}\triangledown_{\nu}-
g_{\mu\nu}\square\big)f_R(R,T)\right].
\end{eqnarray}
We consider the Friedman--Robertson--Walker (FRW) space--time described by the the metric
\begin{eqnarray}\label{13}
ds^2=dt^2-{a(t)}^2\big(\frac{dr^2}{1-kr^2}+r^2d\Omega^2\big),
\end{eqnarray}
where $a(t)$ is the scale factor and $k$ denotes the spatial curvature. Thus, the $00$ and $ii$ components of the field equation (\ref{10}) can be written as
\begin{eqnarray}\label{14}
3(H^2+\frac{k}{a^2})=\rho^{(ef\mbox{}f)}+\rho^{(curv)},
\end{eqnarray}
\begin{eqnarray}\label{15}
(2\dot{H}+3H^2+\frac{k}{a^2})=P^{(ef\mbox{}f)}+P^{(curv)},
\end{eqnarray}
where $\rho^{(ef\mbox{}f)}$, $\rho^{(curv)}$, $P^{(ef\mbox{}f)}$ and $P^{(curv)}$ respectively are
\begin{eqnarray}\label{16}
\rho^{(ef\mbox{}f)}=\frac{1}{f_R(R,T)}\left[(1+f_T(R,T))\rho_m+p_m\right],
\end{eqnarray}
\begin{eqnarray}\label{17}
\rho^{(curv)}=\frac{1}{f_R(R,T)}\left[\frac{1}{2}\big(f(R,T)-Rf_R(R,T)\big)-3H\big(\dot{R}f_{RR}(R,T)-\dot{T}f_{RT}(R,T)\big)\right],
\end{eqnarray}
\begin{eqnarray}\label{18}
P^{(ef\mbox{}f)}=-\frac{1}{f_R(R,T)}p_m,
\end{eqnarray}
\begin{eqnarray}\label{19}
P^{(curv)}=\frac{1}{f_{R}}\left[\frac{1}{2}(Rf_R-f)+2H(\dot{R}f_{RR}+\dot{T}f_{RT})+\ddot{R}f_{RR}
 +\dot{R}^2f_{RRR}+2\dot{R}\dot{T}f_{RRT}+\ddot{T}f_{RT}+\dot{T}^2f_{RTT}\right].
\end{eqnarray}
The equation of state parameter of curvature is then obtained as
\begin{eqnarray}\label{20}
\omega^{(curv)}=\frac{P^{(curv)}}{\rho^{(curv)}}=-1+\frac{\ddot{R}f_{RR}+\dot{R}f_{RRR}+2\dot{R}\dot{T}f_{RR}+\ddot{T}^2f_{RTT}-
H(\dot{R}f_{RR}+\dot{T}f_{RT})}{\frac{1}{2}(f-Rf_R)-3H(\dot{R}f_{RR}+\dot{T}f_{RT})}.
\end{eqnarray}


\section{Particle creation in $f(R,T)$ gravity}
The concept of particle production in an expanding universe has been studied by several authors \cite{parker,zeldovich,agrib,grish,barrow,pavlov}. The study of particle creation in a modified gravity necessarily requires one
to consider quantization in curved background \cite{brill,grib,mukhan}.
However, in curved space-time the Poincar\'{e} group symmetry is not obeyed, then the definition of particles and vacuum states become ambiguous.
There are two different methods of quantization in curved background.
In the first method, the vacuum is described as an entity for which the lowest--energy mode goes to zero smoothly
in the far future and past, represented by "out" and "in" modes with a definite number of particle \cite{brill,fulling},
and the second method is diagonalizing of instantaneous Hamiltonian by a Bogoliubov transformation, which leads to finite
result for the number of created particles \cite{agrib,pavlov}. Since any modified gravity theory such as $f(R,T)$ or $f(R)$ is a recast of general relativity with some non--minimal coupling \cite{lobo}, thus in order to derive the rate of particle production we use non--minimal coupled theories of gravity. For this
purpose, we discuss the Bogoliubov transformation in the context of homogeneous and isotropic cosmology. In this work, we consider the case of $f(R,T)$ gravity
(the case $f(R)$ gravity has been discussed in \cite{salvatore}).

\quad To construct of the Bogoliubov transformation we restrict our attention to the case of a spatially flat FRW metric, and a de-Sitter space--time with a constant curvature $R_0$. So, the metric may be written as
\begin{eqnarray}\label{}
ds^2=dt^2-a^2(t)d\textbf{x}^2=a^2(\eta)(d{\eta}^2-d\textbf{x}^2),
\end{eqnarray}
where $t$ is the cosmic time, $\eta=-\frac{1}{Ha(t)}$ is the conformal time and $\textbf{x}\equiv(x_1,x_2,x_3)$.
In the coordinates $(\eta , \textbf{x})$, and by choosing $a\phi(\eta , \textbf{x})=\chi(\eta , \textbf{x})$, the action is given by \cite{winitzki}
\begin{eqnarray}\label{31}
S=\frac{1}{2}\int d^{^{3}}\textbf{x}d\eta\left[\chi'^2-(\nabla\chi)^2-(m^2a^2+(6\xi-1)\frac{a''}{a})\chi^2\right],
\end{eqnarray}
where the prime denotes derivatives with respect to $\eta$. The equation of motion for $\chi(\eta , \textbf{x})$ is obtained
\begin{eqnarray}\label{32}
\chi''-\nabla^2\chi+(m^2a^2+(6\xi-1)\frac {a''}{a})\chi=0.
\end{eqnarray}
By expanding the field $\chi$ in Fourier modes as
\begin{eqnarray}\label{33}
\chi(\eta , \textbf{x})=\int \frac{d^{^{3}}\textbf{k}}{(2\pi)^{\frac{3}{2}}}\chi_k(\eta)e^{i(\textbf{k}.\textbf{x})},
\end{eqnarray}
and by using the wave vector $\textbf{k}\equiv (k_1 , k_2 , k_3)$, the equation of motion can be written as
\begin{eqnarray}\label{34}
\chi''_k(\eta)+\omega^2(\eta , k  ,\xi)\chi_k(\eta)=0,
\end{eqnarray}
which is analogues to the time--dependent harmonic oscillator with frequency
\begin{eqnarray}\label{35}
\omega_k(\eta)=\left[k^2+a^2(m^2+2f(\xi)H^2)\right]^{\frac{1}{2}},
\end{eqnarray}
where $f(\xi)\equiv6\xi-1$, $H\equiv\frac{\dot{a}}{a}=\frac{a'}{a^2}$, and $\frac{a''}{a}=2a^2H^2=\frac{2}{\eta^2}$.
For $\xi>6$, the function $f(\xi)$ is positive--definite and we can define an effective mass $M_{eff}$ as
\begin{eqnarray}\label{36}
\frac{M_{eff}^2}{H^2}=\frac{m^2}{H^2}+2f(\xi).
\end{eqnarray}
The frequency $\omega_k(\eta)$ is given by
\begin{eqnarray}\label{37}
\omega_k(\eta)=\sqrt{ k^2+\frac{M_{eff}^2}{H^2\eta^2}}.
\end{eqnarray}
The general solution of equation (\ref {34}) is \cite{pinco}
\begin{eqnarray}\label{38}
\chi_k(\eta)=\sqrt {\eta}(A_{k}H^{(1)}_{k,\nu}(\eta)+B_{k}H^{(2)}_{k,\nu}(\eta),
\end{eqnarray}
where $H^{(1)}_{k,\nu}(\eta)$ and $H^{(2)}_{k,\nu}(\eta)$ are Hankel's function of first and second kind, respectively. To calculate the particle production rate, we compare particle number at early and late time. For $\eta\rightarrow 0^{-}$, we obtain
\begin{eqnarray}\label{39}
\chi_k(\eta)\rightarrow\frac{\sqrt |\eta|}{\pi\nu}\big\{\sin(\pi\nu)\Gamma(1-\nu)(\frac{k\eta}{2})^\nu-\Gamma(1+\nu)(\frac{k\nu}{2})^{-\nu}\big\},
\end{eqnarray}
where $\nu\equiv\sqrt{\frac{1}{4}-\frac{M_{eff}^2}{H^2}}$. By calculating the Bogoliubov coefficients, one obtains the number of particle created
 in the $k$ mode as follows \cite{kahn,brander}
 \begin{eqnarray}\label{40}
 |\beta_k|^2=\frac{\omega_k}{2}|\phi_k(\eta)-\frac{i}{\omega_k(\eta)}\dot{\phi}_k(\eta)|^2.
 \end{eqnarray}
 In the case of $\frac{ M_{eff}}{H}\gg1$, we obtain
 \begin{eqnarray}\label{41}
 |\beta_k|^2\sim\frac{H^3}{32{\pi}m^3}|\Gamma(1-\frac{im}{H})|^2\exp(\pi m),
\end{eqnarray}
which can give rise to an accelerating expansion phase. In the case $\frac{M_{eff}}{H}\ll1$, we will have
\begin{eqnarray}\label{42}
|\beta_k|^2\ll 1.
\end{eqnarray}
In this case, we have two different states. The first is $\frac{m}{H}\ll 1$, with $m\rightarrow 0$.
The second is un--physical, because the Bogoliubov coefficient $\beta_k$ is diverging \cite{kolb}.
In a semi classical procedure the Bogoliubov coefficients allow to pass from a vacuum state to another one, and different classes of $f(R,T)$
gravity change the vacuum states according to Bogoliubov transformation. Therefore the Bogoliubov coefficients strictly depend
on the form of $f(R,T)$ gravity, and minimizing the Bogoliubov coefficients
correspondingly minimize the rate of particle production. In doing so, we can fix the free parameters of a given $f(R,T)$ model.

\section {Minimizing the particle production rate in $f(R,T)$ gravity}

 Our strategy to minimize the rate of particle production and then constrain the free parameters of a given model, is to use de-Sitter phase \cite {congola} {\it i.e.} $R=R_0$ and the condition $T=T_0$. By taking the trace of field equation (\ref{9})  and setting $T_{\mu\nu}=0$, we obtain
\begin{eqnarray}\label{43}
R_0f_R(R_0 , T_0)-2f(R_0 , T_0)=4P_mf_T(R_0 , T_0).
\end{eqnarray}
We can define an effective cosmological term as
\begin{eqnarray}\label{44}
\Lambda_{eff}=\frac{f(R_0 , T_0)}{2f_R(R_0 , T_0)}+\frac{f_T(R_0 , T_0)}{f_R(R_0 , T_0)}=\frac{R_0}{4}.
\end{eqnarray}
For the physical case $\frac{m}{H}\ll1$ in the context of $f(R,T)$ gravity, the Taylor expansion of equation (\ref{41}) at first order is given by
\begin{eqnarray}\label{45}
|\beta_k|^2=e^{\pi m}\left[\frac{H^3}{32\pi m^3}+\gamma^2\frac{H}{32\pi m}\right],
\end{eqnarray}
where we use $\Gamma(1-\frac{im}{H})\sim1+i\gamma\frac{m}{H}$, for $\frac{m}{H}\ll1$ and $\Gamma$, $\gamma$  are Euler
function and Euler constant($\gamma\sim0.577$), respectively.
Besides, in a flat FRW universe from equation (\ref{14}) we can write
\begin{eqnarray}\label{46}
H^2=\frac{1}{3}(\rho^{(eff)}+\rho^{(curv)}).
\end{eqnarray}
For the case of $T=T_0$ and $R=R_0$, we obtain
\begin{eqnarray}\label{47}
 |\beta_k|^2=e^{\pi m}\left[\frac{1}{3^{3/2}32\pi m^3}\big(\frac{(1+f_{0T})\rho_0}{f_{0R}}-
 \Lambda_{eff}\big)^{3/2}+\frac{\gamma^2}{32{\sqrt3}\pi m}\big(\frac{(1+f_{0T})\rho_0}{f_{0R}}-\Lambda_{eff}\big)^{1/2}\right],
 \end{eqnarray}
 where $f_{0T}\equiv f_T(R=R_0, T=T_0)$, $f_{0R}\equiv f_R(R=R_0, T=T_0)$, $f_0\equiv f(R=R_0, T=T_0)$ and $\rho_0$ is the value of standard matter--energy
 density for $R_0$, $T_0$.\\
 Since the form of $f(R,T)$ is not known {\it a priori}, by evaluation of the different vacuum states for some classes of $f(R,T)$ function according to Bogoliubov
 transformations, we can minimize the rate of particle production. In doing
 so, we can fix the free parameters of a given $f(R,T)$ model.
 Some interesting models of $f(R,T)$ gravity are
 \begin{eqnarray}\label{48.a}
 f_1(R,T) &=R+2\lambda T,
 \end{eqnarray}
 \begin{eqnarray}\label{48.b}
 f_2(R,T) &=R+\varepsilon R^2-T+T^2,
 \end{eqnarray}
 \begin{eqnarray}\label{48.c}
 f_3(R,T)=\alpha R^n+\beta T^{-m},
 \end{eqnarray}
 \begin{eqnarray}\label{48.d}
 f_4(R,T)=R+R^n+\delta R^{-m}+\sigma T^\ell.
 \end{eqnarray}
 The coefficients $\lambda, \varepsilon, \alpha, \beta, n, m, \delta$ and $\sigma$ can be fixed by equation (\ref{47}).
 In order to do this, we require the rate of particle production to be negligible or essentially as small as possible \cite{peebles}.

 Now, we assume that the free parameters of equations (\ref{48.a})-(\ref{48.d}) minimize the Bogoliubov coefficients. This procedure is performed by taking derivatives of $|\beta_k|^2$ with respect to the parameters of the models and imposing
them to be equal to zero. The results are summarized in the following Table:\\

 \begin{tabular}[b]{|l|c|c|}
 \hline
$ f_n(R,T)$ & num.param. & minimize.\\
 \hline \hline \hline
 $f_1(R,T)$ & 1 & $1+2\lambda\geq \frac{R_0}{4\rho_0}$\\
 \hline
 $f_2(R.T)$ & 1 & $\varepsilon\leq \frac{4\rho_0T_0}{R_0^{2}}-\frac{1}{2R_0}$ \\
 \hline
 $f_3(R,T)$ & 4 & $R_0(n\alpha R_0^{n-1}-1)+4\beta m\rho_0T_0^{-m-1}\leq 4\rho_0-R_0$ \\
 \hline
 $f_4(R,T)$ & 5 & $nR_0^n(1-\frac{m}{n}\delta R_0^{-m-n})-4\ell \sigma \rho_0 T_0^{\ell-1}\leq 4\rho_0-R_0$ \\
 \hline
  \end{tabular}

\vspace{5mm}
\begin{center}
{\footnotesize{ Table 1: }}\hspace{-2mm} {\scriptsize Table of minimizing condition for the free parameters of $f(R,T)$ models. \\The equality sign corresponds to the state of vanishing Bogoliubov coefficients, whereas the inequality sign corresponds to the state, where the Bogoliubov coefficients are not zero.}\\
\end{center}
\quad The free parameters in this Table, namely $\rho_0, R_0, T_0$, can be set by using late--time and CMBR cosmological constraints \cite{farooq,ade}.
Then, the values of model parameters, namely $\lambda, \epsilon, \alpha, \beta, \delta, \sigma, n, m$, are constrained through the inequalities mentioned
in Table 1 for each $f(R,T)$ model. By any choice of consistent modified $f(R,T)$ gravity, we find the following general inequality
\begin{eqnarray}\label{52}
\frac{f_R(R,T)}{1+f_T(R,T)}\leq\frac{4\rho_0}{R_0},
\end{eqnarray}
which for the special case of $f(R)$ modified gravity reduces to the same
result obtained in \cite{salvatore}
\begin{eqnarray}\label{53}
{f_R(R)}\leq\frac{4\rho_0}{R_0}.
\end{eqnarray}

\section{Conclusion}
We have considered the role of particle production rate in the context of $f(R,T$) gravity. In this regard, we have
obtained the Bogoliubov coefficients which provides a way to pass from a vacuum state to another one. These coefficients can be
connected to the Hubble parameter which strongly depends on the functional form of a given $f(R,T$) model. Thus,
the particle production rate is related to the specific form of $f(R,T$) gravity. So, this is a  procedure to constrain the free
parameters of the model, invoking a semiclassical scheme. We have assumed
that the Bogoliubov coefficients can be minimized in passing through different vacuum states, once the background is postulated. Particularly, we have studied the de-Sitter phase $R = R_0$, $T=T_0$ and derived
the Bogoliubov transformations for some typical forms of $f(R,T$) models. The Bogoliubov coefficients have been evaluated for a homogeneous and isotropic universe, assuming that the particle production rate is negligibly small. This provided us with the conditions on the free parameters of
$f(R,T$) models which have been reported in Table.1.

\end{document}